\documentclass[floatfix, onocolumn,nofootinbib,longbibliography,preprintnumbers,10pt]{revtex4}

\usepackage[T1]{fontenc}
\usepackage{palatino}

\usepackage{braket} 

\usepackage{graphicx}

\usepackage{amssymb,amsmath,amsthm}

\usepackage{bm} 
\usepackage{graphicx} 
\usepackage{multirow} 
\usepackage{cases} 

\usepackage{booktabs,tabu,fancybox}

\usepackage{verbatim}  

\usepackage{xspace}
\makeatletter
\DeclareRobustCommand\onedot{\futurelet\@let@token\@onedot}
\def\@onedot{\ifx\@let@token.\else.\null\fi\xspace}

\def\eg{\emph{e.g}\onedot} 
\def\ie{\emph{i.e}\onedot} 
 
\def\etc{\emph{etc}\onedot}

\def\viz{\emph{viz}\onedot}
\makeatother

\usepackage{etoolbox} 
\newtoggle{omitauthor}
\newtoggle{omitwordcount}
\toggletrue{omitwordcount}
\newtoggle{omitdraftnumber}
\toggletrue{omitdraftnumber}
\newtoggle{omitdraftfootnote}
\toggletrue{omitdraftfootnote}

\usepackage[shortlabels,inline]{enumitem}  
\newlist{inlinelist}{enumerate*}{1}  
\setlist[inlinelist]{label=(\roman*)}  
\usepackage{hyperref}

\usepackage{caption}
\usepackage{subcaption}  

\usepackage{chngcntr}

\newcommand{\vect}[1] {\ensuremath{\mathbf{#1}}} 

\newcommand{\oplabel}[1]{\ensuremath{\mathcal{#1}}} 

\newcommand{\LB}{London-Bauer\ }
\newcommand{\MW}{Many Worlds\ }
\newcommand{\dBB}{de Broglie--Bohm\ }

\newcommand{\MWns}{Many Worlds}
\newcommand{\dBBns}{de Broglie--Bohm}

\begin{document}


\title{Systematizing the Interpretation of Quantum Theory via Reconstruction\iftoggle{omitdraftfootnote}{}{\footnote{This is a draft.  Comments, suggestions, and critical feedback are most welcome.}}}

\iftoggle{omitauthor}{}
{\author{Philip Goyal}	
    \email{pgoyal@albany.edu}
    \affiliation{University at Albany~(SUNY), NY, USA}
}
\date{\today}
\iftoggle{omitauthor}{}{\homepage[Homepage:~]{https://www.philipgoyal.org}}
\linespread{1.418}

\begin{abstract}

For a century, quantum theory has posed a fundamental challenge to philosophical thinking.  On its face, it repudiates many of the key features of the mechanical conception of physical reality.  However, the challenge of developing a precise, coherent alternative to that conception has yet to be met.    Here, I argue that a major hindrance to the project of quantum interpretation is its existing interpretative methodologies, which suffer from a lack of systematicity in their judgements about what aspects of the theory are interpretational relevant.  In particular, I argue that current interpretations tend to marginalize the informal part of the theory in favour of its formal part, and place inappropriate emphasis on the natural language component of the formalism over its detailed mathematical structure.  To counterbalance these biases, I propose that an interpretation-free zone be constructed around the theory, wherein an interpreter initially adopt a descriptive stance which considers all parts of the theory, and that the results of this deliberation~(and the judgements about what facts are interpretationally relevant) are reported as part of their interpretation.

I argue that the interpretation of quantum theory poses special challenges and difficulties which necessitate this interpretation-free zone, and that existing interpretative methodologies are insufficient to address them.  Further, I argue that a reconstructive interpretative methodology, which harnesses the recent results of the quantum reconstruction program, provides a powerful means to identify almost all facts that could be interpretationally relevant, and naturally meets these challenges and difficulties.  Moreover, I argue that the quantum reconstruction program offers a powerful way to discover new physical principles, and offers a systematic pathway to build a rich, coherent conception of quantum reality.

\end{abstract}
\linespread{1.618} 

\iftoggle{omitdraftnumber}{}{\preprint{Draft, v1.00}}

\maketitle


\section{Introduction}

The quantum reconstruction program seeks to obtain a deeper understanding of quantum theory by deriving the quantum formalism from physically well-motivated postulates and assumptions formulated in an operational framework~\cite{Rovelli96,Hardy01a,Fuchs2003,Grinbaum-reconstruction-interpretation,Grinbaum-reconstruction, Grinbaum03,Goyal2022c}.  In a nutshell, its goal is to \emph{do for quantum theory what Einstein did for the Lorentz transformations}---to systematically trace the mathematical structure of quantum theory~(\eg its complex vector space structure) to physical principles and concepts that are precisely articulated within an operational framework, thereby distilling the mathematical structure of quantum theory into a few physical postulates and assumptions whose operational meaning is unambiguous.   

The reconstruction program came to prominence in the 1990s and early 2000s~\cite{Rovelli96, Hardy01a, Fuchs2003}, inspired in part by \begin{inlinelist} \item the striking insights and associated technological discoveries of the quantum information revolution in the 1980s and 1990s~(\eg~\cite{BB84, Bennett1993}); \item Wheeler's arguments for an informational perspective~(`It from Bit') on physical theory~(see~\cite{Wheeler89} and also \eg~\cite{Grinbaum03,Timpson2013,Goyal12}); and \item the insights afforded by no-go theorems, such as Bell's theorem~\cite{Bell64}, which were formulated in an operational setting. \end{inlinelist} The early successes of the quantum reconstruction program---most notably Hardy's reconstruction of the finite-dimensional von Neumann--Dirac axioms~\cite{Hardy01a}---established the viability of the program's aspirations.  In the subsequent twenty-five years, most components of the quantum formalism have been reconstructed; in particular:~\begin{inlinelist} \item the finite-dimensional von Neumann--Dirac axioms, from a wide variety of angles~(\eg~\cite{Hardy01a,Hardy01b, Goyal-QT2c,GKS-PRA, DakicBrukner2010, Dariano-operational-axioms,Chiribella2011, Goyal2014, Hohn2017, MullerMasanes2016, MasanesMullerAugusiakPerez-Garcial2013,SelbyScandoloCoecke2018}); \item the quantum wave equations~(e.g.~\cite{DarianoPerinotti2014,Goyal-correspondence-rules-2010}); \item the identical particle formalism~\cite{Goyal2015,Goyal2019a,SanchezDakic2024}; and \item the classical-quantum correspondence rules~\cite{Goyal-correspondence-rules-2010}. \end{inlinelist}

The successes of the quantum reconstruction program naturally raise the question of the \emph{relationship} between the quantum reconstruction and interpretation programs.  In the early days of quantum reconstruction, outspoken physicist advocates positioned reconstruction as an \emph{alternative} to traditional interpretation~\cite{Rovelli96,Fuchs2003}.  Some philosophers of physics who engaged with reconstruction in this early phase reflected this view~(e.g.~\cite{Grinbaum-reconstruction-interpretation,Grinbaum-reconstruction, Grinbaum03}), while others cast doubt on the explanatory value of reconstruction~\cite{Brown-Timpson2006}.  However, in the last few years, a number of more nuanced reappraisals of the relationship between the two programs have appeared~\cite{Berghofer2023, Oughton2022, Goyal2022c,Oddan2025}.  For example,~\cite{Berghofer2023} systematically analyses several possible relationships between reconstruction and interpretation, and concludes that \emph{quantum reconstructions are themselves in need of interpretation,} while~\cite{Goyal2022c} proposes a \emph{reconstructive approach to interpretation}.  In addition, several works in recent years examine the reconstructive methodology more broadly---its historical use in the elucidation of various physical theories~\cite{Darrigol2015b}, its relation to axiomatic methods in physics and mathematics~\cite{Oddan2024}, its explanatory function~\cite{Dickson2015,Felline2016,Oughton2022,Oddan2025}, and its relevance to the phenomenology of quantum physics~\cite{BerghoferGoyalWiltsche2019,IslamiWiltsche2025}.    Nevertheless, despite the wealth of reconstructive results, it remains the case that, in the context of interpretative discussions, the reconstruction program is rarely mentioned~\cite[fn.~1]{Berghofer2024}; and, if mentioned, is still sometimes opposed to interpretation~(e.g.~\cite[Ch.~10]{French2023}).

The goal of this paper is \begin{inlinelist} \item to present a \emph{reconstruction-based methodology} for the interpretation of quantum theory; \item to provide an understanding of the \emph{kind of interpretations} that reconstructions enable; and \item to provide an understanding of the \emph{specific benefits} of reconstruction-fuelled interpretations. \end{inlinelist} 
The reconstruction-based interpretative methodology is based around four key ideas:
\begin{enumerate}
\item\emph{Interpret both the formal and informal parts of quantum theory.}  The practice of quantum physics involves a set of theoretical ideas and tools, some of which are exact, others that are heuristic in nature.  It also involves intricate experimental practices.  Physics provides empirical warrant for the \emph{entire} practice, and any facet of this practice could be of potential interpretative relevance.  Hence, interpret the practice \emph{as a whole}, rather than, say, focussing~(as is customary) on the quantum theoretic formalism.

\item \emph{Establish an interpretation-free zone.}  Judgements concerning what parts of quantum practice are interpretatively relevant ought to be decoupled from the act of interpretation.  This is essential in order to protect against pervasive interpretational biases, such as the tendency to accord primary interpretative value to crisp formalism over heuristics or experimental practices.  Hence, in the first instance, create an \emph{interpretation-free zone} around quantum theory.  Within this zone, adopt a \emph{descriptive}---rather than an \emph{explanatory} or \emph{interpretative}---stance which does not favour any part of the practice.

\item \emph{Employ reconstruction to pinpoint what is interpretationally relevant.}  Reconstructing any part of the quantum formalism is a highly non-trivial task.  Therefore, if a reconstruction is successful, one can be confident that \emph{it has distilled much~(if not all)  that is theoretically and experimentally relevant} for interpretation of that part of the formalism.  Hence, consult the available reconstructions to assist in pinpointing the interpretationally relevant parts of the practice of quantum theory.

\item\emph{Situate quantum theory within a coherent metaphysical conception.}  The broad philosophical aims of interpretation require that one understand how quantum theory differs from classical physics.  Hence, situate quantum theory within a metaphysical conception of reality that can be systematically compared with the mechano-geometric-atomistic conception of classical physics.
\end{enumerate}
We show that, in comparison to the prevailing interpretative methodologies, a \emph{reconstruction-based interpretational methodology} confers several benefits, including: \begin{enumerate}[(i)] \item Exposing more of the physical content implicit in the mathematical structure of the quantum formalism to philosophical reflection; \item Bracketing the metaphysical connotations of both the natural-language and symbolic components of the quantum formalism; \item Cleanly separating that part of the quantum formalism which is empirically warranted from justificatory or interpretative language that is customarily associated with the formalism;  \item Surfacing interpretatively-relevant assumptions that are implicit in experimental practice~(in both experimental design and in the processing of bare experimental data);  \item Rendering somewhat vague concepts~(such as complementarity and latencies) analytically precise; and \item Facilitating the discovery of new physical principles of interpretative import.  \end{enumerate}

The paper is organized as follows.  Sec.~\ref{sec:interpretation-of-physical-theories} delineates the parts of the practice associated with a physical theory which need to be considered for interpretational purposes, examines the interpretational biases that tend to arise in the interpretational process, and proposes an interpretation-free zone as a safeguard against these biases.  Sec.~\ref{sec:reconstruction} explicates how the methodology of reconstruction elucidates physical theory via \emph{stratification} and \emph{operationalization}, and how it assists in identifying facts of potential interpretational relevance.  This is followed by a brief overview of the goals of the quantum reconstruction program, and its results to date.  Sec.~\ref{sec:interpretation-of-quantum-theory} describes the challenges and special difficulties which are faced when attempting to interpret quantum theory, and analyses existing interpretative methods with respect to how they handle these challenges and difficulties.  Finally, Sec.~\ref{sec:reconstructive-interpretation} considers the reconstructive approach to quantum interpretation, and summarize some recent insights it has yielded.  Sec.~\ref{sec:conclusion} contains a summary and outlook.

\section{Interpretation of Physical Theories}
\label{sec:interpretation-of-physical-theories}

To appreciate the implications of the quantum reconstruction program for the interpretation of quantum theory, it is necessary to first consider what a physical theory---broadly construed---consists of; and to investigate the interpretational biases that are usually associated with the various components of a theory.

\subsection{Components of a Physical Theory}
\label{sec:components-of-a-physical-theory}

It is commonplace for a physical theory to be identified with its core formalism, \emph{viz.} its physical principles, laws, postulates, and axioms.  For example, if one were asked to give a summary of Newton's theory of mechanics, one would probably write down various principles~(such as Galileo's principle of relativity, and the Principle of Inertia), together with Newton's laws of motion, the parallelogram of forces, and Newton's law of gravitation.  These would be expressed in a mixture of mathematical symbols and idealized, technical language, such as \emph{space}, \emph{time}, \emph{body}, \emph{mass}, and \emph{force}.

However, as is well known, the application of the idealized formalism of a theory to the physical world requires that an interface be established between \emph{idealized thought}, on the one hand, and \emph{sense}, on the other.  The part of the theory which facilitates this interface can be divided into two parts:~
\begin{enumerate}[label=\arabic*.]
\item \emph{Experimental practices.} That which guides the design of experiments which allow measurements of position, acceleration, mass, etc.; and which specifies how to interpret the raw data generated by such experiments.
\item \emph{Modelling heuristics.} That which enables the delineation of `physical systems' of interest, and the construction of \emph{theoretical models} of such systems.
 \end{enumerate} 
This part of the theory is relatively \emph{informal}---it is largely learned through intensive problem solving in the classroom, and during research apprenticeship in a theory group or in a laboratory.  In other words, it is part of a physicists' \emph{know-how}.  It is far more often \emph{illustrated} or \emph{shown} through carefully-selected examples rather than explicitly articulated.  And, if articulated, it takes the form of heuristics which are context-dependent rules-of-thumb that, unlike the theory's formalism, make no claim to generality.     
\begin{figure}
\begin{center}
\includegraphics[width=\textwidth]{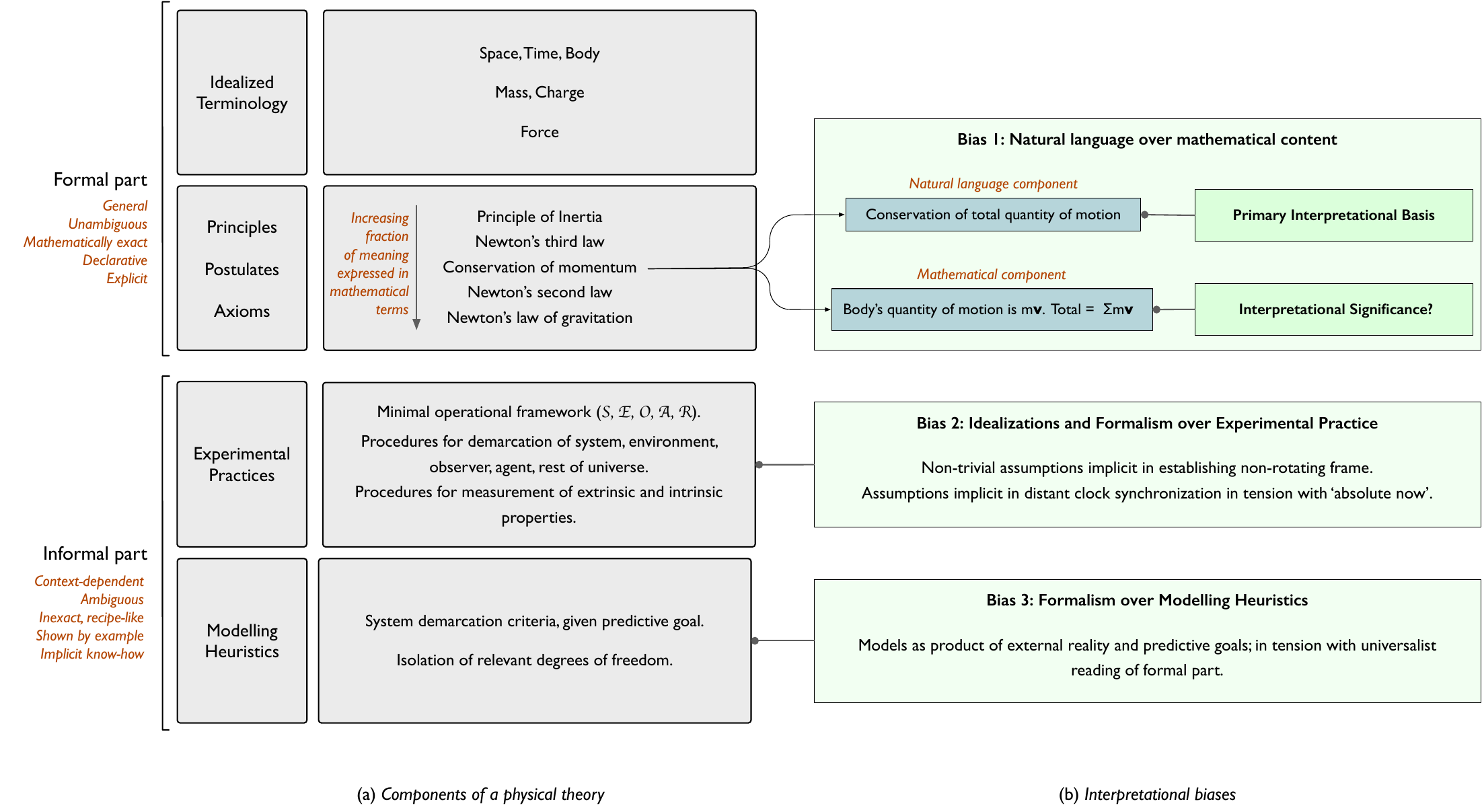}
\caption{\label{fig:theory-and-interpretational-biases}\emph{(a)~Components of a physical theory~(for interpretational purposes), illustrated using classical point mechanics. (b)~Types of interpretational bias, with examples drawn from text.}}
\end{center}
\end{figure}

\subsection{Interpretational Biases}
\label{sec:interpretational-biases}

The fact that a physical theory is regarded as empirically adequate in some domain of physical phenomena depends on both the formal and informal parts of the theory~(see Fig.~\ref{fig:theory-and-interpretational-biases}(a)).  Hence, on the assumption that one regards a physical theory worthy of philosophical investigation because it is an \emph{empirically validated system of thought}, as opposed to being merely a mental construction disconnected from sensory experience, \emph{any part of either the formal or informal parts of the theory could be of interpretative relevance.} 

However, interpreters~(be they physicists or philosophers of physics) typically make up-front judgements about what part of a theory they deem interpretationally significant, and thereby introduce significant bias into their interpretations at the outset.  These biases closely track the different components of a theory described above.  Below I consider three of the main interpretational biases~(see summary in Fig.~\ref{fig:theory-and-interpretational-biases}(b)), and investigate their consequences.

\subsubsection{Bias 1:  Natural language over mathematical details.}
\label{sec:bias1}
As mentioned above, a theory's formalism is expressed in a mixture of mathematical symbolism and natural language.  The `fraction' of physical content reflected in natural language versus mathematical symbolism varies considerably with the type of component of the formalism.  For example, in a theory's \emph{principles}, a significant fraction, or sometimes even all, of the meaning is carried by natural language.  A few principles, such as the principle of inertia, are entirely intelligible without recourse to mathematical symbolism.  However, such principles are rare, and lie at one extreme of a spectrum---the content of most principles, such as the principle of conservation of momentum, is partially contained in their mathematical component.  

Typically, as one moves from principles to \emph{laws}, \emph{postulates}, and finally to \emph{axioms}, progressively less of the physical content is reflected in the natural language component.  For example, the content of Newton's second law is largely encapsulated in the equation~$\vect{F} = m\vect{a}$.  The axioms of quantum theory~(the von Neumann--Dirac axioms) lie at the other extreme of the spectrum---very little of their physical content can be `read off' their natural language components.

\emph{The first bias is the tendency to place interpretational emphasis on the natural language component of a theory's formalism, or on specific \emph{consequences} of the formalism which can be expressed in natural language.}  If the physical meaning of the mathematical components of the formalism seems obscure, it tends to be interpretationally marginalized, despite its essential role in the theory's empirical adequacy.  Two examples below drawn from classical physics illustrate this bias and its consequences.  Quantum theory will be considered separately in Sec.~\ref{sec:interpretation-of-quantum-theory}.

\begin{figure}
\begin{center}
\includegraphics[width=\textwidth]{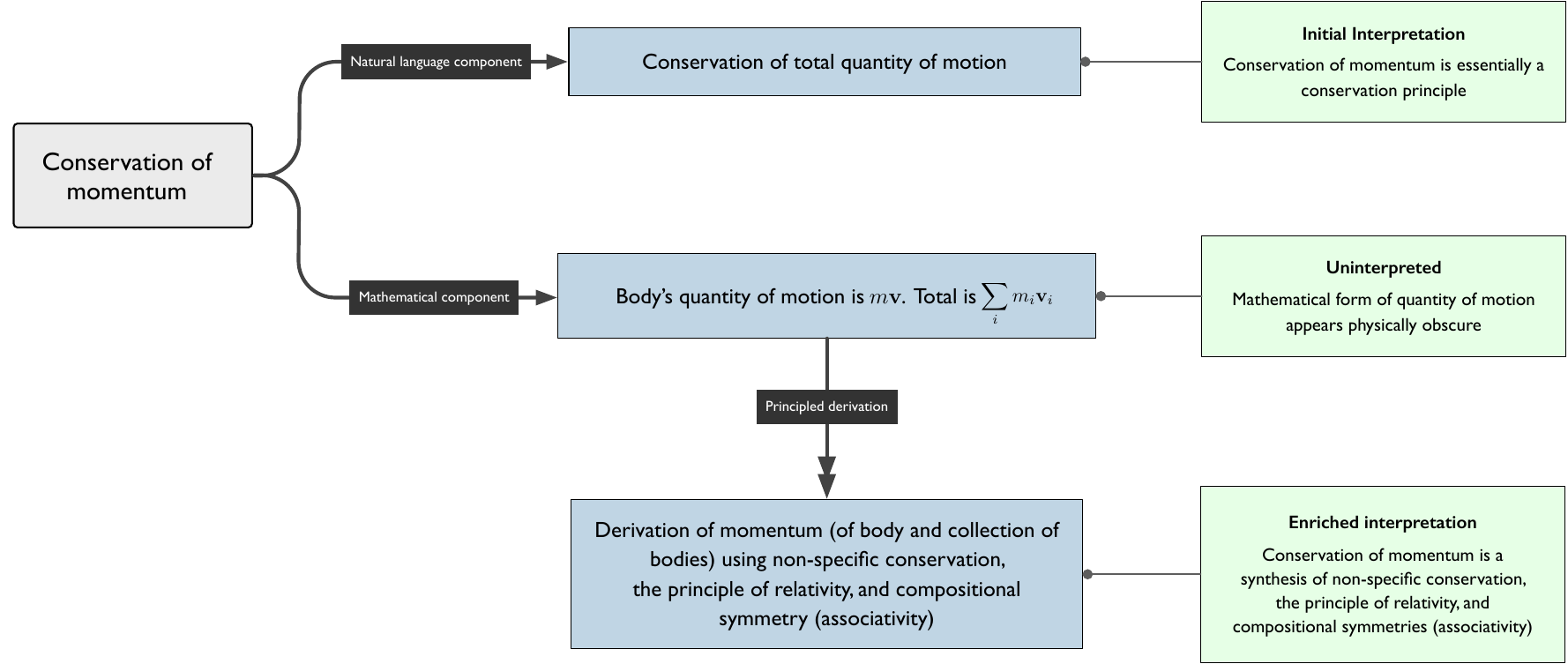}
\caption{\label{fig:conservation-of-momentum}\emph{Interpretation of Conservation of Momentum.}}
\end{center}
\end{figure}

\paragraph{Example 1:~Conservation of momentum.} \label{sec:conservation-of-momentum}
Consider the classical mechanical statement that the total momentum,~$\sum_i m_i \vect{v}_i$, of a set of isolated, interacting bodies is conserved under their dynamical evolution.  This statement is the conjunction of two distinct statements~(see Fig.~\ref{fig:conservation-of-momentum}).
\begin{enumerate}[label=(\arabic*)]
\item \emph{Non-specific conservation.} The \emph{total quantity of motion} of an isolated system of interacting bodies remains constant under dynamical evolution.
\item \emph{Quantity of motion.} The quantity of motion of a body~$m$ with velocity~$\vect{v}$ is~$m\vect{v}$; and the total quantity of motion of a set of bodies is~$\sum_i m_i \vect{v}_i$.
\end{enumerate} 
The \emph{physical meaning} of the original statement is commonly taken to be that given by the first of these statements, \emph{viz.} the natural language component of the statement.  For this reason, the statement is usually simply described as a `conservation principle'.

\begin{figure}
\begin{center}
\includegraphics[width=\textwidth]{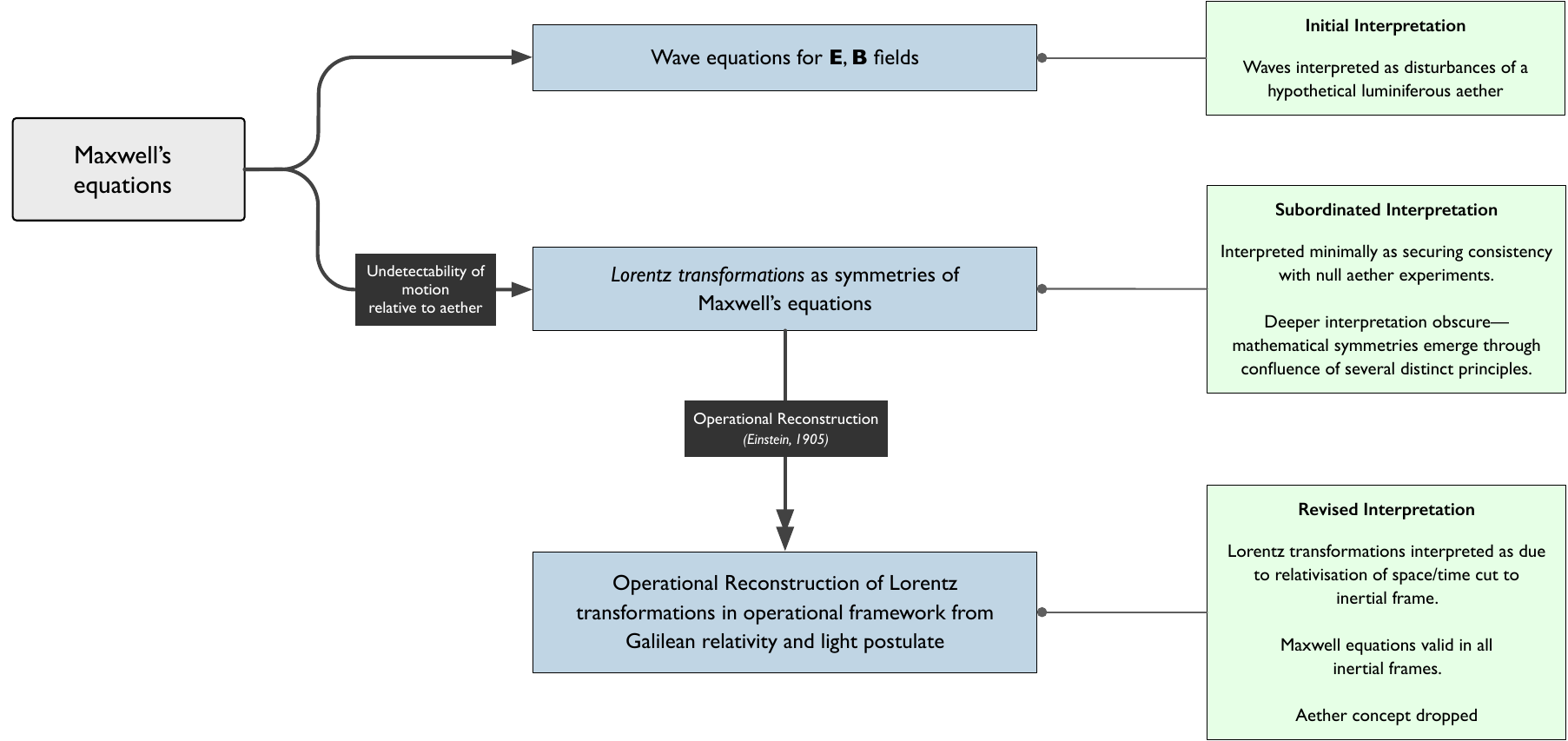}
\caption{\label{fig:Maxwell}\emph{Interpretation of Maxwell's equations.}  }
\end{center}
\end{figure}

Nevertheless, there exist principled derivations of momentum~(and total momentum)~(e.g.~\cite{Schutz1897,Tolman1912,Ehlers-Rindler-Penrose1965,Desloge1976a,Desloge1976b,Goyal2020}).  It is striking that these only began to appear two centuries after the introduction of the vectorial momentum. %
The most compelling of these derivations show that a combination of non-specific conservation and Galileo's principle of relativity, together with compositional symmetries, yield the mathematical form of momentum and total momentum.  Hence, the mathematical form of the quantity of motion embodies a wealth of additional physical content.  Thus, taking into account extant insight into the physical origin of this form considerably enriches the initial interpretation of the conservation of momentum based on its natural-language component~(see also Sec.~\ref{sec:stratification-operationalization} and Fig.~\ref{fig:classical-physics-stratification-momentum}). %
\paragraph{Example 2:~Maxwell's equations.} \label{sec:Maxwells-equations}
Following the discovery that Maxwell's equations support electromagnetic waves, it was almost universally supposed that there exists a medium, the \emph{luminiferous aether}, in which these waves propagate~(see Fig.~\ref{fig:Maxwell}).  
This inference was based on an analogy with then-known mechanical waves, such as sound waves, which could be successfully modelled as disturbances propagated through a medium.  The conviction that electromagnetic waves were disturbances of a medium had sufficient force that the aether hypothesis was upheld even decades after the failure of experiments, most famously Michelson--Morley's 1887 experiment, to detect motion relative to the putative aether.  For example, in 1906, two decades after Michelson--Morley's experiment, Poincar\'e proposed a principle of relativity which posited the impossibility of detection of motion relative to the aether~\cite{Poincare1906}.  

The requirement of undetectability of motion relative to the aether culminated in work of Lorentz~\cite{Lorentz1904} and Poincar\'e~\cite{Poincare1906} with the formulation of a symmetry of Maxwell's theory consisting of the Lorentz transformations together with corresponding transformations of~$\vect{E}, \vect{B}$, on the assumption that charge is invariant.  But the deeper physical meaning of the Lorentz transformations was opaque---they were not the direct expression of a physical principle, but emerged circuitously from the confluence of the several distinct physical principles~(Gauss' law, Faraday's law of induction, the non-existence of magnetic monopoles, Amp\`ere's law, and Maxwell's displacement current) encoded in Maxwell's equations, together with the empirically-forced requirement embodied in Poincar\'e's relativity principle.  As a result, their interpretation was subordinated to that inferred from a natural-language consequence of Maxwell's equations~(\emph{viz.} the existence of a luminiferous aether inferred from the existence of electromagnetic waves).

Yet, in hindsight, the interpretative significance of a mathematical symmetry of Maxwell's equations~(which was needed to explain a striking and unexpected empirical fact, \emph{viz.} the non-detection of motion relative to the putative aether) was under-appreciated.  Einstein's derivation of the Lorentz transformations on the basis of simple physical postulates~(the Galilean principle of relativity and the independence of the one-way speed of light from the motion of the source) within an operational framework~(invoking a natural time-synchronization protocol---see Sec.~\ref{sec:clock-synchronization}) provided a clear physical interpretation these transformations.  This lead to the abandonment of the natural-language-based interpretation, and to the construction of a new physical interpretation in which the Lorentz transformations were viewed as a general statement of the physical transformation between inertial frames.  Maxwell's equations were accordingly \emph{reinterpreted} as valid in \emph{every} inertial frame, and electromagnetic waves were thenceforth regarded as a new kind of wave requiring no medium.  

As a result of this process, an initial interpretation, which projected an \emph{existing} physical understanding of the wave concept onto the newly-discovered electromagnetic waves, was replaced by an interpretation that eschewed that projection, and moreover prompted a radical revision~(see Fig.~\ref{fig:classical-physics-stratification-einstein}) of multiple fundamental concepts~(space/time cut relativised to inertial frame; the relation of matter to energy; etc.) which penetrated through to the \emph{metaphysical foundations}~(the \emph{categories} of Fig.~\ref{fig:classical-physics}) of classical physics.  However, this interpretation only arose through creative, reconstructive engagement with a mathematical structure emerging from Maxwell's equations whose physical meaning was initially obscure.

\subsubsection{Bias 2:  Theory over Experimental practice.}
\label{sec:bias2}

Application of the formal part of a theory to reality as sensorily experienced requires an \emph{interface} between \begin{inlinelist} \item the idealized theoretical notions which are employed in the formal part of the theory; and \item the concrete sensory experiences of an experimenter. \end{inlinelist}   This interface consists of two parts:
\begin{enumerate}
\item\emph{Operational framework.} A conceptual framework which specifies \begin{inlinelist} \item the minimum distinctions that must be made; and \item the capabilities that must exist \end{inlinelist}
 in the sensory realm in order that the theory can have empirical traction.
\item\emph{Experimental practices.} A set of practices which specify in detail \emph{how} the distinctions and capabilities specified in the operational framework are practically realized.
\end{enumerate} 
\emph{The second bias in the interpretation of a physical theory is the tendency to presume that the operational framework and the associated experimental practices, which together empirically ground the theory, are less interpretationally relevant than the formal part of the theory.}  However, the operational framework and experimental practices rest upon assumptions which, once made explicit, are seen to be both numerous and non-trivial\footnote{The theory ladenness of experimental practices is well known~(\eg \cite{Kuhn1996,Hanson1958}). However, making the theoretical assumptions explicit, so that they can be examined in a critical light, is a challenging task.  The penetrating analyses of Helmholtz~\cite{Helmholtz1977} and Poincar\'e~\cite{Poincare1898}~(see Sec.~\ref{sec:clock-synchronization}) on space and time measurement are exemplars of making such long-standing assumptions explicit.  Assumptions implicit in the measurement of time-independent properties of quantum particles are discussed in~\cite{Goyal2023a}.}. Moreover, they are sometimes in tension with key idealized notions of the theory.  Indeed, the surfacing of such assumptions has played a critical role in the development of many physical theories, including special and general relativity, and quantum mechanics.  Hence, such assumptions are critical to take into account during the interpretational process.  Below we first summarize the key features of the operational framework of a theory, and briefly illustrate the associated experimental practices.  We then give an example from classical physics which exemplifies the impact on interpretation of this bias.   Examples from quantum theory will be considered in Sec.~\ref{sec:interpretation-of-quantum-theory}.

\paragraph{Operational framework.} \label{sec:operational-framework}
At minimum, that which is concretely experienced in the sensory realm must be able to partitioned into \emph{system}~$\oplabel{S}$ and \emph{non-system}~$\overline{\oplabel{S}}$. The latter must, in turn, be able to be partitioned into the \emph{environment},~$\oplabel{E}$, \emph{observer},~$\oplabel{O}$, \emph{agent},~$\oplabel{A}$, and \emph{rest of the universe},~$\oplabel{R}$.

Here, \emph{environment} refers to that part of the physical world which is \emph{directly relevant} to the behaviour of the system in the context of the experiment, while the \emph{rest of the universe} contributes to its behaviour indirectly.  For example, the overall form of the dynamical laws that govern~$\oplabel{S}$ are presumably a reflection of the relationship between~$\oplabel{S}$ and~$\oplabel{R}$.

The \emph{observer} is that which is capable of observation but not action, while the \emph{agent} is that which is capable of \emph{controllably manipulating} the environment.  The observer and agent is often embodied in the same entity, \emph{viz.} the \emph{experimenter}, but this need not be the case.   In addition, the observer can observe and the agent can act without being influenced in what they choose to observe, or what they choose to do, by the system itself.  This is a no-conspiracy assumption.

Furthermore, an \emph{experiment} takes place over an interval of time.  Hence, an operational notion of \emph{time} common to all entities in the experiment~(except the rest of the universe), for the duration of the experiment, is necessary.  For the entirety of that duration, the above partitioning must hold fast.

To the above \emph{minimal operational framework} must be added operations that \emph{measure} the properties that are theoretically ascribed to objects.  For example, there must exist operations for measurement of distances and durations, and for measurement of static properties such as mass and charge.  These operations will, in general, be dependent upon the theory in question.

\paragraph{Experimental practices.}
The \emph{substantiation} of the operational framework's distinctions and capabilities is made possible by a body of experimental practices.  These stipulate how the above partitioning is to be carried out, how a common time is to be established, and how properties are to be measured.  These practices are an outgrowth of our experiences in everyday life and pre-scientific practices, and are refined as science develops.  As previously mentioned, they are generally learned by apprenticeship, by showing and doing rather than through explicit description.

Implicit in these practices are numerous non-trivial assumptions which are carried over from everyday life, pre-scientific practices, or practices associated with earlier theories.  For instance, an elementary procedure to measure `the length' of an object employs a standard unit-length object, which is then moved alongside the object to `mark out' its length.  For this procedure to work, it must be assumed that this unit-length is invariant under displacement.  Length comparison in different directions requires that one also assume that the unit-length is invariant under rotation.  Similar assumptions are implicit in procedures for measurement of adjacent durations~(\eg measured by adjacent clocks).  The comparison of durations at different locations requires additional assumptions. Experimental procedures on quantum systems require yet further non-trivial assumptions.  For example, measurement of the intrinsic properties~(such as mass and charge) of quantum objects rests upon assumptions of these objects' persistence and reidentifiability~\cite{Goyal2023a}.

\paragraph{Example:~Poincar\'e and Einstein on clock synchronization.}
\label{sec:clock-synchronization}
The metaphysical foundations of Newtonian mechanics rest upon an absolute partitioning of space and time.  Time is posited to flow equably in all places.  An ideal clock at any location is presumed to register this flow of time.  Consequently, two widely-separated clocks beat at the same rate.  This empirically grounds the idealized notion of `absolute now'.

However, following Mach's operational critique~\cite{Mach1919} of Newtonian mechanics, the assumptions implicit in our actual experimental procedures for duration-comparison were subjected to critical analysis by Poincar\'e~~\cite{Poincare1898,Poincare1902}.  This analysis exposed a subtlety:~suppose we have a procedure for synchronizing two adjacent clocks; then any conceivable procedure for the synchronization of the clocks \emph{once separated} necessarily invokes an \emph{additional} assumption.  Although the exact form of this additional assumption will depend upon the selected procedure, that an additional assumption is required is unavoidable.  Yet, we have no independent way to confirm the correctness of this assumption.  Hence, according to Poincar\'e, the additional assumption amounts to a \emph{conventional choice} that is ultimately justified by the relative simplicity brought by this choice to our theoretical description of nature.  Consequently, there is no accessible empirical counterpart to the notion of `absolute now'---\emph{there is a unbridgeable gap between a fundamental theoretical notion~(`absolute now') and what can be objectively empirically established via experimental procedures.}

To elaborate:~first suppose that we agree that if an embodied observer hears two immediately adjacent clocks tick simultaneously over a given duration, then they are synchronized for the observed duration.  How can we establish synchrony of distant clocks?  Due to the localized, embodied nature of observers, no observer can be present at both clocks at the same time.  Hence, we have two choices:
\begin{enumerate} 
\item\emph{Transport of initially-synchronized clocks.}  Synchronize two clocks locally, and then separate them widely.
\item\emph{Signal-based synchronization.} Start with two widely-separated clocks, and then use a signal to synchronize them.
\end{enumerate}
Both procedures rest upon non-trivial physical assumptions.  The first assumes that the clocks' synchrony is unbroken by transport.  The second that the signal's speed is independent of direction.  However, we have no independent empirical means to check the validity of either assumption.  In particular, using a single clock, we can only measure the round-trip speed of light, not the one-way speed.

As is well-known, Poincar\'e's penetrating operational analysis was pivotal in Einstein's reconstruction of the Lorentz transformations.  In~\cite[\S1]{Einstein05}, Einstein presents an operational framework containing a signal-based clock synchronization protocol in which a light signal is transmitted from clock~$A$ to clock~$B$ and then immediately back to clock~$A$.  Denoting by~$t_1, t_3$ the times of transmission and receipt at clock~$A$, and~$t'_2$ the transmission and receipt at clock~$B$, the clocks are said to be synchronized if~$t'_2 = (t_1 + t_3)/2$. Combined with Einstein's light postulate, the relativization of simultaneity to an inertial frame follows at once.  With the addition of a convention for comparison of lengths in different inertial frames, the Lorentz transformations follow.

The broader lesson here~(which we will apply to quantum theory in Sec.~\ref{sec:interpretation-of-quantum-theory}) is as follows.  Critical analysis of experimental procedures can expose decisive gaps between idealized theoretical notions and what can be empirically established or measured.  The awareness of such gaps can be pivotal to the interpretation of existing physical theories, or indeed the creation of new ones.

\subsubsection{Bias 3:  Formalism over modelling heuristics.}
\label{sec:bias3}

The result of applying the \emph{general} formalism of a theory to a \emph{particular} physical system of interest is a \emph{physical model} of that system.  However, the model does not capture the entirety of the actual physical system.  Rather, it describes the physical system relative to a particular predictive goal.  

One fundamental role of modelling heuristics is to provide criteria for demarcating a system relative to a particular goal.  This is necessary because, even assuming that one has reliable procedures to demarcate that portion of the universe which can be successfully described by, say, Newtonian mechanics, the existence of long-range gravitational interactions means that there is no \emph{natural} way to to carve up this portion into entirely \emph{isolated} subsystems.  For example, if ones goal is to model the movements of the planets in the solar system, one can regard this as `a system' for modelling purposes only if ones measurements are below a certain level of precision.  Beyond that level, `the system' must be enlarged to include planetoids, the solar wind, and so on.

Additional criteria are needed to isolate certain degrees of freedom of a given physical object.  For example, from a mechanical viewpoint, any non-idealized extended body has a vast number of degrees of freedom.  Usually, one is only interested in a few of these.  Isolation of these degrees of freedom requires heuristics, which enable the modeller to determine under what conditions these degrees of freedom will remain effectively isolated from the body's other degrees of freedom.

\emph{The third bias in the interpretation of a physical theory is the tendency to presume that modelling heuristics, which are essential for the construction of empirically viable models, are less interpretationally relevant than the formal part of the theory, or not relevant at all.}  This can manifest as the implicit identification of a theoretical model of a system with the physical system itself.  It also often manifests as the \emph{universalist} assumption that one can take the system to be \emph{any} part of the universe~(irrespective of ones predictive goal), or that the theory applies to the universe as a whole.  The latter leads to claims such as the universe is mechanical~(in the context of classical mechanics), or that one can take any physical system at any scale to be a quantum system.  However, a theory such as Newtonian mechanics or quantum theory has only been precisely empirically validated in very restricted circumstances.   Hence, any such interpretative assumption is an extrapolation of the theory into an empirically untested~(and possibly in principle untestable) regime.  But, on the previously stated assumption that a physical theory galvanises philosophical interest precisely because it is an empirically battle-tested system of thought, it is essential to take into account its domain of established validity and avoid such unwarranted extrapolations.

In addition, the heuristics associated with a theory may embody non-trivial assumptions about the nature of the physical world which are in tension with key idealized notions of the theory.   And, even if one has concluded that the heuristics associated with, say, Newtonian mechanics are not interpretationally relevant, one cannot simply infer that the heuristics associated with quantum theory are also interpretationally irrelevant.  For example, a quantum model of a system invariably depends upon a classical description of the system.  For instance, a quantum model of the electrons in a bar of silver depends upon the geometry of the bar.  This theoretic dependency of quantum theory upon classical physics for the construction of models is striking, and may have interpretational implications.

\subsection{Factors Behind Interpretational Biases}
\label{sec:factors-behind-interpretational-biases}

As we have seen above~(\S\ref{sec:interpretational-biases}), the interpretation of physical theories~(or parts thereof) is subject to a number of evidential selection biases.  These biases reflect many different factors, including historical, disciplinary, and cognitive.  

\begin{enumerate}[1.,leftmargin=2.5em]
\item\emph{Historical factor.}
When first created, a physical theory commonly contains mathematical features whose physical rationale is obscure.  In the first phase of life of a theory, these features are primarily justified \emph{post hoc}, on the grounds that the theory \emph{works}.  A deeper physical understanding of these features is typically not developed until decades~(or even centuries) afterwards\footnote{\label{fn:delay-reconstruction}The reasons for this long delay are several:~\begin{inlinelist} \item the development of such understanding usually requires looking at the theory from a radically new angle; \item new or unfamiliar mathematics is required to prove the new perspective; \item the existence of a workable theory disincentivises search for a deeper understanding. \end{inlinelist} }~\cite{Darrigol2015b,Goyal2022c}.  However, the physical and metaphysical understanding of a theory tends to be developed during the first phase, which tends to be based on whatever aspects of a theory can be rendered into natural language at the time.  This understanding then becomes canonical, and is reflected in textbooks and the research literature.  

Hence, when a deeper understanding of these mathematical features is eventually developed, the revision of the physical and metaphysical understanding of a theory must contend with a widely-shared, long-sedimented understanding that is incomplete or at odds with the new developments.  And due to the widespread canonical understanding, it is quite possible that a would-be interpreter of a theory~(whether a physicist or philosopher of physics) will be unaware of the existence of this deeper understanding of the mathematical features.

\item\emph{Disciplinary factor.} Highly intellectualized disciplines, such as physics and philosophy, prize knowledge that is explicit, declarative, exact, and general.  For example, physics strives to discover and precisely formalize the most general patterns that govern the particular.  Although philosophy is more diverse in its attitudes to know-how and implicit knowledge, philosophy of physics is presently dominated by the analytical view, which emphasizes the explicit and declarative.  This results in a general disciplinary bias towards the formal part of a theory, away from the implicit knowledge and know-how encapsulated in modelling heuristics and experimental practice.

\item\emph{Cognitive factor.} Engagement with the theories of physics requires a certain mathematical and analytical sophistication. Mathematical and analytic thinkers will naturally tend to gravitate towards a theory's mathematical formalism as opposed to its relatively informal and context-dependent heuristics and know-how.  Contrariwise, more holistic thinkers will naturally pay more attention to the latter, but may  have a weaker understanding of the former.
\end{enumerate}

\subsection{Establishing an Interpretation-free Zone}

To recap, the line of thinking in Sec.~\ref{sec:interpretation-of-physical-theories} thus far is as follows.  First, a physical theory is an empirically-validated system of thought.  Such empirical validation requires a set of experimental practices and modelling heuristics, which interface the theory's formalism with empirical reality.  Second, insofar as an empirically-validated physical theory is deemed of interpretational interest, any part of its formal or informal parts could be of interpretational significance.  Third, there are pervasive interpretational biases which lead interpreters to focus on certain parts of a theory and to marginalize other parts---\viz \begin{inlinelist} \item to focus on those aspects of the formalism~(or specific consequences thereof) which can be readily expressed in natural language, and to marginalize those mathematical forms whose physical meaning seems obscure; \item to marginalize the operational framework, associated experimental practices, and modelling heuristics, in favour of the formal part of the theory.  \end{inlinelist}

Furthermore, these interpretational biases are rooted in factors that are largely beyond the control of the individual interpreter~(see Sec.~\ref{sec:factors-behind-interpretational-biases}).  Therefore, special measures are needed to manage them.  Accordingly, we make the following general recommendation:~prior to beginning any interpretative project, the interpreter adopt a \emph{descriptive} stance towards the theory of interest.  That is, prior to engaging in interpretation \emph{per se}, they carry out a careful appraisal of all the conceivably relevant formal and informal aspects of a theory.  This should include \begin{inlinelist} \item a thorough study of any attempts to unravel the physical meaning of the theoretical formalism; \item a study of the assumptions implicit in the operational framework and associated experimental procedures; and \item a study of the modelling heuristics. \end{inlinelist}   And the interpreter ought to systematically describe their appraisal when reporting on their interpretation, so that what evidence has and has not been judged interpretationally relevant~(and the justification for those judgements) is explicit.

This descriptive stance creates an \emph{interpretation-free zone} around the physical theory of interest, which enhances the \emph{objectivity} of the interpretational process.  And since the descriptive stance counters the superficial discarding of conflicting evidence, or the pre-emptive dismissal of evidence which may be lacking mathematical precision, formality, and generality, it will tend to lead to richer, more provocative interpretations.   

The proposed interpretation-free zone is analogous to the safe-zones which have been painstakingly established in experimental science in order to better insulate data collection and analysis from a wide range of biases.  For example, in drug trials, the well-known safeguard of double-blinding ensures that patients and clinicians are kept in the dark about whether a given patient is in the control group or treatment group.  In triple-blinded experiments, those responsible for data analysis are sometimes added to this list.  However, even such safeguards have been found wanting.  In recent years, the so-called replication crisis~\cite{OpenScienceCollaboration2015} has lead to calls for the strengthening of best practices.  For example, in fields as diverse as psychology and particle physics, pre-registration of experiments~(\emph{e.g.} to counter non-reporting of null results), pre-registration of data analysis procedures prior to data collection~(\emph{e.g.} to counter hypothesizing after the results are known~(HARKing)~\cite{Kerr1998}), and making all raw data and methods of analysis publicly available~(\emph{e.g.} to counter manipulation of data or analyses) are increasingly common, albeit not yet widespread.

The descriptive stance advocated here is characteristic of the phenomenological method of philosophical enquiry~(e.g.~\cite[\S1.2.10]{BerghoferWiltsche2019}; see also~\cite{BerghoferGoyalWiltsche2019, Bitbol2023}).  For example, in practicing a Husserlian epoch\'e (or~\emph{bracketing}) of the so-called `natural attitude', one strives to suspend ones pre-reflective belief in the idea that there exist physical objects with a mind-independent reality.  The interpretation-free zone advocated above seeks to bracket not only pre-existing metaphysical beliefs or predilections, but a variety of deep-seating biases arising from a range of historical, disciplinary, and cognitive factors.

\section{Reconstruction of a Physical Theory}
\label{sec:reconstruction}

A descriptive stance helps to create an interpretation-free zone around a theory.  However, as the examples of interpretational bias in Sec.~\ref{sec:interpretation-of-physical-theories} illustrate, it is exceptionally difficult to judge \emph{a priori} which facts are interpretationally relevant.  Moreover, as we have seen, interpretationally-relevant facts are often encoded in mathematical formalism or implicit in experimental practices, and their decoding or explication requires intricate mathematical and/or conceptual analysis.  Hence, we need specific methods in order to pinpoint and precisely articulate those facts that could be interpretationally relevant.

As seen in the aforementioned examples, a recurring motif is that of deriving or reconstructing a mathematical component of a theory's formalism in order to understand its deeper physical meaning.  This method has been used repeatedly in the history of physics to elucidate the physical meaning of the mathematical formalism of physical theories~\cite{Darrigol2015b}.   Einstein's reconstruction of the Lorentz transformations is the farthest reaching example of this methodology in action, for two reasons.  First, it incorporates both an incisive analysis of experimental practices~(distant-clock synchronization) and a principled derivation of a physically-obscure emergent mathematical structure~(the Lorentz transformations).  Second, it leads not only to a new \emph{physical} interpretation of Maxwell's equations~(\emph{viz.} \begin{inlinelist} \item these equations are valid in every inertial frame; and \item electromagnetic waves need no medium\end{inlinelist}) but to a re-evaluation of categorical \emph{metaphysical} distinctions fundamental to classical physics~(absolute space/time cut, \etc).

\subsection{Stratification and Operationalization}
\label{sec:stratification-operationalization}
\begin{figure}
\begin{center}
\includegraphics[width=\textwidth]{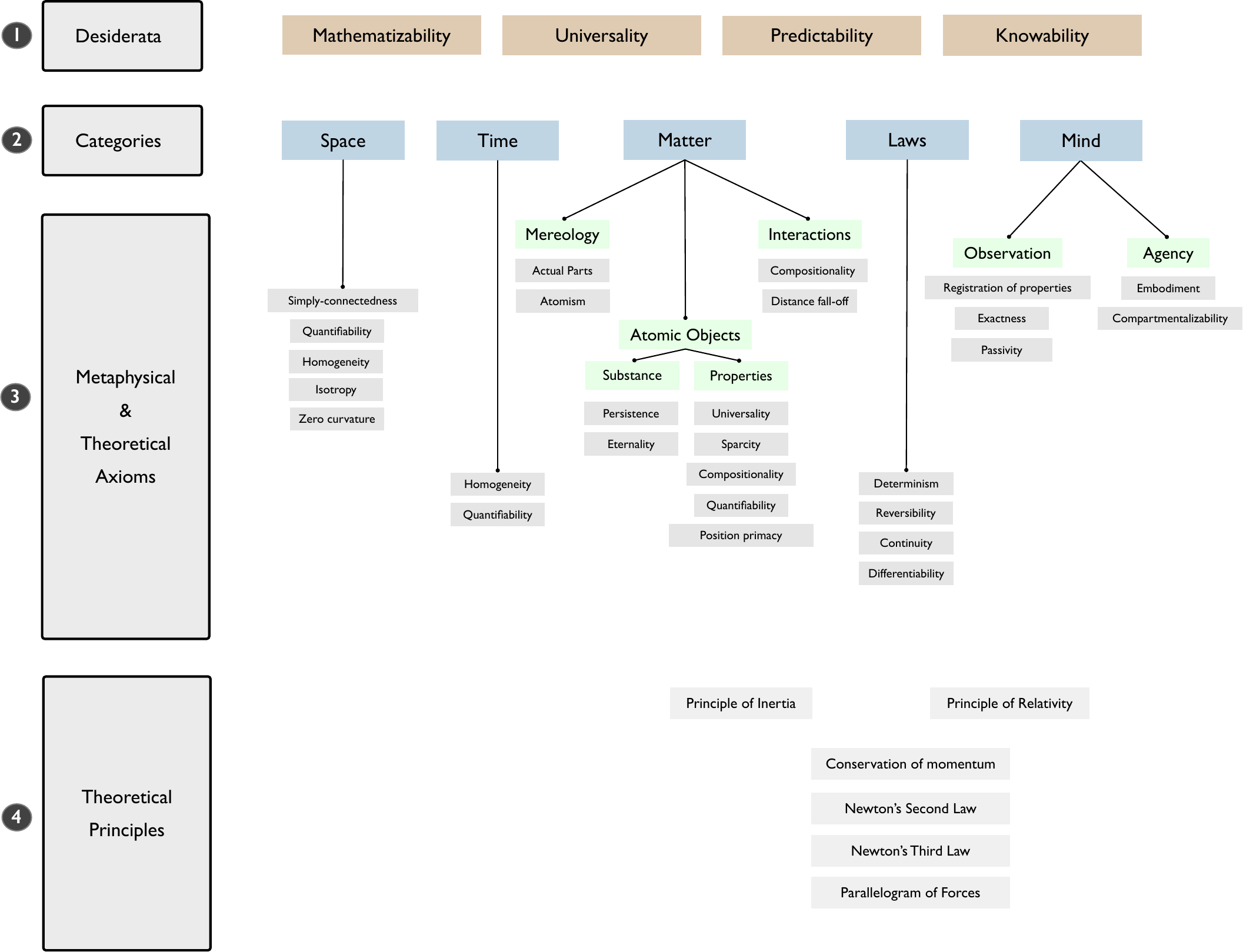}
\caption{\label{fig:classical-physics}\emph{Schematic overview of classical physics..} A depiction of the classical conception of physical reality prior to Maxwellian electrodynamics, based on~\cite[Sec.~II]{Goyal2022c}.  This hierarchical conception is divided into several levels: 1.~\emph{Overarching desiderata}, which shape the entire conception.  2.~\emph{Categories}~(space, time, matter, laws, mind); 3.~\emph{Metaphysical and physical axioms}; and 4.~\emph{Physical principles} which underpin Newtonian particle mechanics.}
\end{center}
\end{figure}

As the aforementioned examples illustrate, the reconstruction of the formalism of a physical theory~(or a part thereof) elevates our physical and metaphysical understanding through at least two processes:
\begin{enumerate} 
\item\emph{Stratification.}  The mathematical formalism is traced back to \begin{inlinelist} \item existing fundamental physical or metaphysical axioms; \item existing physical principles; or \item newly-formulated metaphysical axioms or physical principles.  \end{inlinelist} 
\item \emph{Operationalization.}  The reconstruction more securely connects the mathematical formalism to the theory's operational framework, sometimes leveraging previously unexamined assumptions implicit in experimental procedures that substantiate the operational framework.
\end{enumerate}
The processes of stratification and operationalization in the above-mentioned examples can be most clearly seen against the backdrop of the overall physical and metaphysical architecture of classical physics.  This architecture is schematically represented in Fig.~\ref{fig:classical-physics}~(based on~\cite[Sec.~II]{Goyal2022c}), and depicts:
\begin{enumerate} 
\item \emph{Desiderata of classical physics.}  Key desiderata---mathematizability, universality, predictability, and knowability---which shape classical physics as a whole.
\item \emph{Categories.}  The fundamental categories---space, time, matter, laws, mind---into which reality is divided.
\item \emph{Metaphysical \& Theoretical axioms.}  Specific metaphysical axioms~(such as the metaphysical doctrine of actual parts~\cite{Holden2006}, which underpins the atomistic conception) and theoretical axioms~(such as isotropy of space), which are broadly shared by the theories of classical physics.
\item \emph{Theoretical principles.} Physical principles and laws belonging to a particular classical theory~(Newtonian mechanics, in this case).
\end{enumerate}

\begin{figure}
\begin{center}
\includegraphics[width=\textwidth]{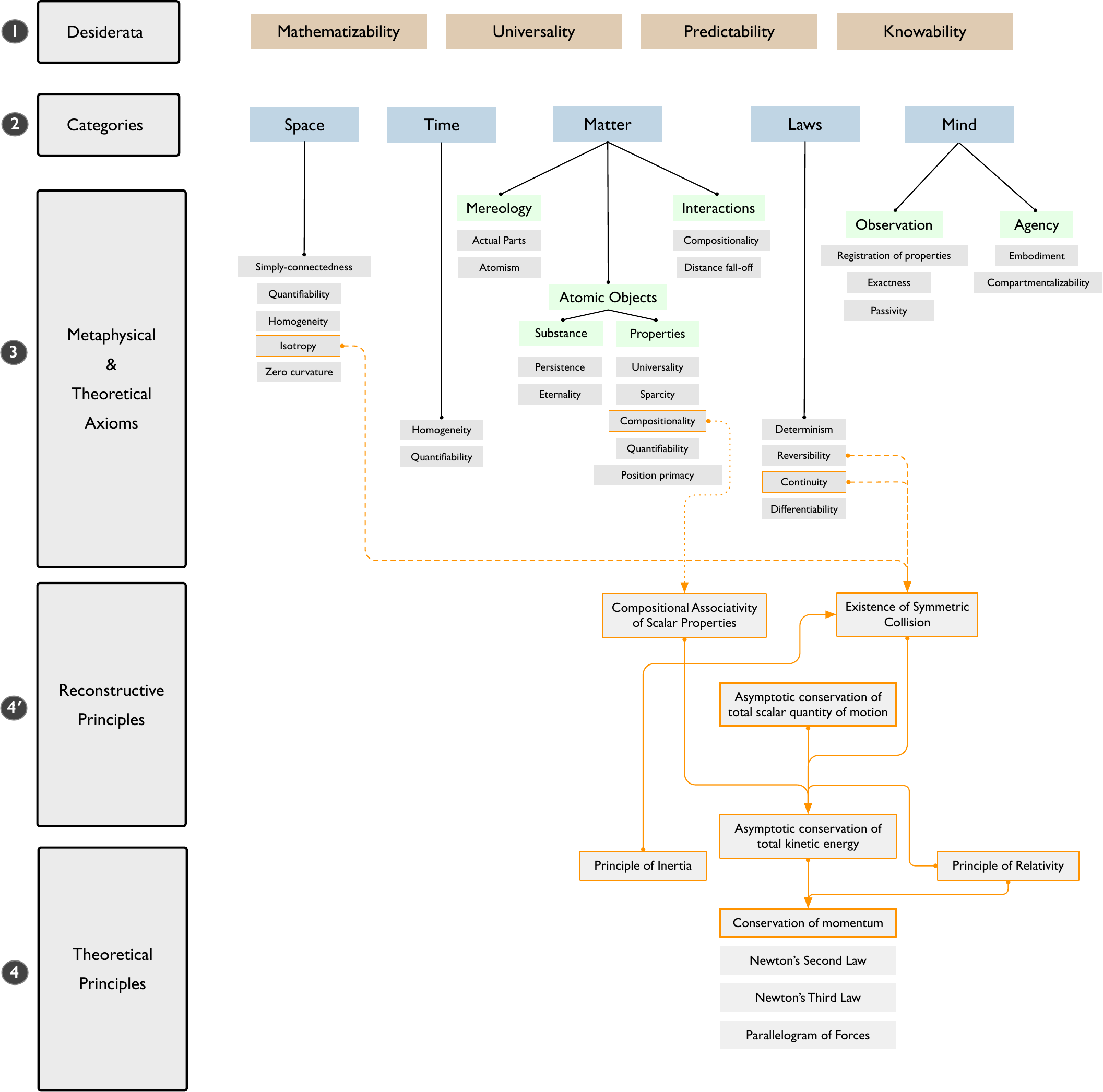}
\caption{\label{fig:classical-physics-stratification-momentum}\emph{Stratification in derivation of conservation of momentum.}  Diagram depicting the stratification achieved by derivation of conservation of momentum~(based on~\cite[\S2]{Goyal2020} and~\cite{Schutz1897}).  A solid line from~$A$ to~$B$ indicates logical dependency of~$B$ on~$A$.  A dashed line from~$A$ to $B$ indicates that $A$~rationalizes~$B$; a dotted line that~$B$ is a precisification of~$A$.} 
\end{center}
\end{figure}

Against this backdrop, the derivation of conservation of momentum~(\S\ref{sec:conservation-of-momentum}) elucidates through stratification.  This is illustrated in Fig.~\ref{fig:classical-physics-stratification-momentum}, using the derivation of momentum contained in~\cite[\S2]{Goyal2020} and~\cite{Schutz1897}. 
In this instance, the reconstruction is based on such assumptions as the \emph{compositional associativity of scalar properties} and the \emph{existence of symmetric collisions.}  These assumptions are more readily intelligible than the conservation of momentum, and thereby serve as \emph{intermediaries} between the metaphysical \& theoretical axioms, on the one hand, and the theoretical principle of the conservation of momentum, on the other.  In particular, as illustrated in the figure, these reconstructive assumptions can be largely justified by reference to elementary metaphysical and theoretical axioms.  And the reconstruction shows how the principle of conservation of momentum~(in its mathematical details, in both the non-relativistic and relativistic instances) can be derived from these assumptions. 

\begin{figure}
\begin{center}
\includegraphics[width=\textwidth]{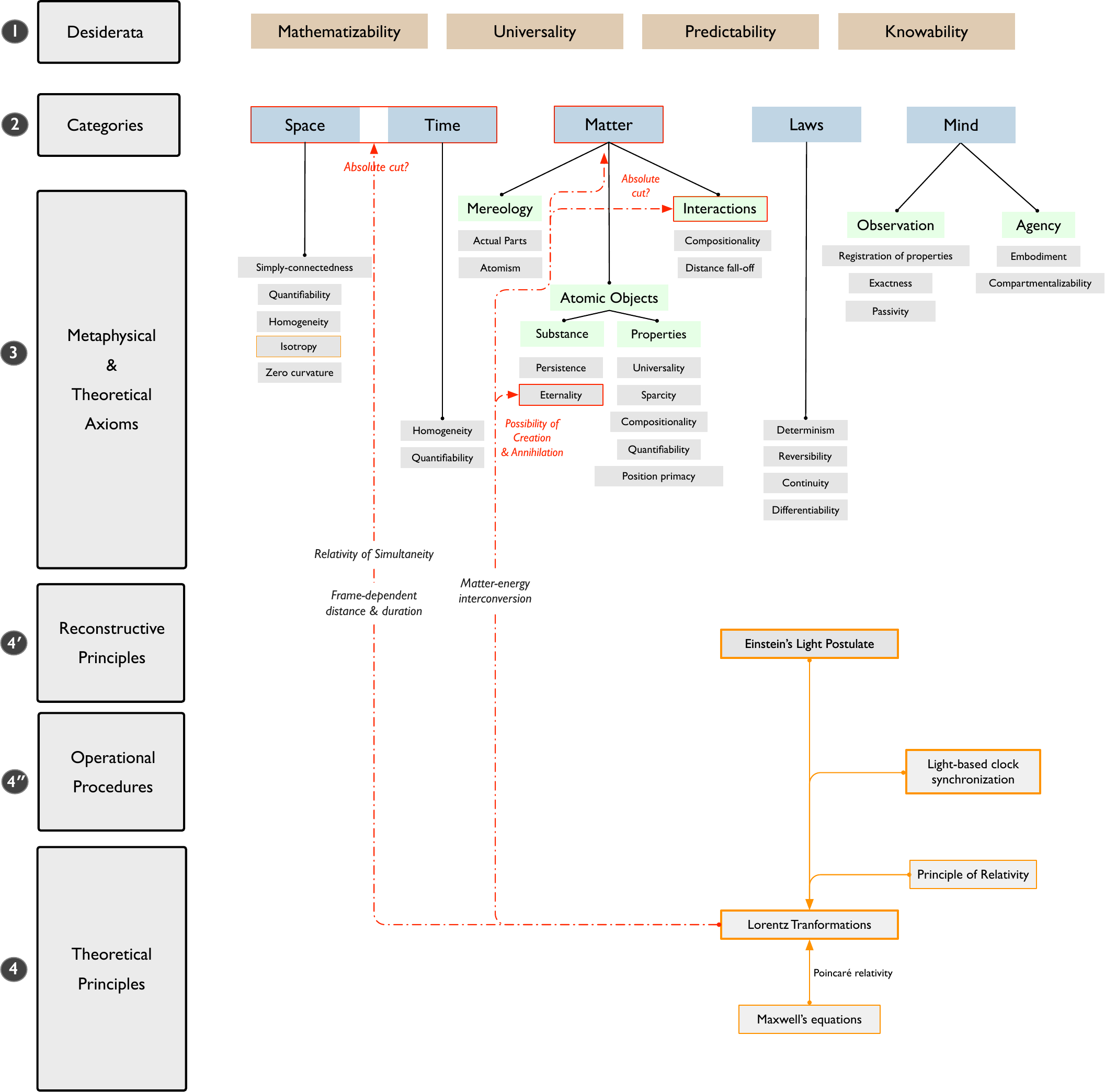}
\caption{\label{fig:classical-physics-stratification-einstein}\emph{Stratification and operationalization in Einstein's reconstruction of Lorentz transformations.}  Diagram depicting Einstein's reconstruction of the Lorentz transformations.  A solid line from~$A$ to~$B$ indicates logical dependency of~$B$ on~$A$.  A dot-dashed line from~$A$ to $B$ indicates that $A$ forces reconsideration of category or (metaphysical, physical) axiom~$B$.  This forcing is either a \emph{direct} forcing~(in case of absolute space/time cut) or an \emph{indirect} forcing via additional reasoning steps~(in case of eternality and the absolute matter/interaction cut).}
\end{center}
\end{figure}

Einstein's reconstruction of the Lorentz transformations~(as elaborated in~\S\ref{sec:Maxwells-equations} and \S\ref{sec:clock-synchronization}) involves both stratification and operationalization.  This is illustrated in Fig.~\ref{fig:classical-physics-stratification-einstein}.
In this case, the reconstruction invokes not only a novel reconstructive principle, namely Einstein's light postulate, but also harnesses Poincar\'e's analysis of distant time measurements.  In this case, the reconstruction casts the Lorentz transformations in a radically new physical light, forcing re-examination of certain metaphysical and theoretical axioms, as well as the fundamental categorization of reality.  For example, the absolute pre-Einsteinian boundary between space and time is brought into question by the relativization of simultaneity to an inertial frame, while matter-energy conversion brings into question the absolute distinction between matter and interactions.

\subsection{Reconstruction of Quantum Theory}
\label{sec:reconstruction-of-quantum-theory}

The goal of the quantum reconstruction program is to articulate a set of \emph{physically well-motivated principles} from which the formalism of quantum theory~(the von Neumann--Dirac axioms plus its auxiliary formalism) can be systematically derived~(e.g.~\cite{Hardy01a,Fuchs2003,Rovelli96}).  

The proponents of the quantum reconstruction program  frequently cite Einstein's reconstruction of the Lorentz transformations as an exemplar.  Two aspects of Einstein's work are particularly noteworthy.  First, as discussed in Sec.~\ref{sec:clock-synchronization}, that he employed an operational framework in which fundamental theoretical notions like `now' are given explicit (albeit idealized) experimental expression. Second, as discussed in Sec.~\ref{sec:Maxwells-equations}, that he articulated assumptions---the principle of relativity and the independence of one-way speed of light from the speed of the source---that could be readily understood, even if their union resulted in highly counter-intuitive consequences.  As a result, he was able to isolate the Lorentz transformations from their context of origin~(Maxwell's theory, and the intricate phenomena of electromagnetism) and reproduce them within an austere operational framework from two readily graspable physical postulates.  This provided the foundation for the broad consensus in physics community to embrace the radical implications~(see Sec.~\ref{sec:stratification-operationalization}) of Einstein's postulates.

Analogously, a reconstructor of quantum theory seeks to systematically derive the formalism of quantum theory within an operational framework which refers to only basic experimental operations~(and possibly harnesses critical analyses of basic experimental procedures), and which is based on physical assumptions that can be readily understood in operational terms.  The goal is to thereby derive the formalism afresh, without reference to the complex physical phenomena~(such as heat capacities of materials and spectroscopic data) which historically inspired the creation of quantum theory, and without recourse to the kind of mathematical guesswork~(such as the use of complex numbers, and Schroedinger's posit that composite systems are described in a higher-dimensional configuration space)~\cite[\S\,III\,A]{Goyal2022c} that was part and parcel of the creative process which historically gave rise to the  quantum formalism.

It is worth pointing out that reconstruction is a creative process which invariably requires discovery of new perspectives very different from the one which gave rise to the theory in the first place, and consequently typically lags theory creation by several decades~(see fn.~\ref{fn:delay-reconstruction} and~\cite[\S1]{Goyal2022c}).  For example, Einstein's reconstruction appeared some 40 years after Maxwell's theory, and made essential use of Poincar\'e operationalist analysis of elementary measurement procedures.  In the case of quantum theory, there were many attempts prior to the 1980s to render quantum theory more intelligible through reconstruction-type work~(\eg \cite{Birkhoff-VNeumann36,Mackey1957,Emch-Piron1963}; see~\cite[\S1]{SelbyScandoloCoecke2018} for an overview of this early work; and~\cite{Mittelstaedt2013} on the related notion of rational reconstruction).  However, these attempts had comparatively little impact, in part due to their frequent recourse to abstract assumptions.  The quantum information revolution in the 1980s inspired the hypothesis that quantum theory might be derivable in terms of information-theoretic principles~\cite{Rovelli96,Fuchs2003}, and lead to the conviction such a derivation could be achieved without recourse to abstract mathematical assumptions that lack clear physical justification~\cite{Hardy01a,Grinbaum-reconstruction, Grinbaum03,Goyal2022c}.  Following promising early information-inspired reconstructive attempts~(e.g.~\cite{Wootters80}), one of the first successful reconstructions of the finite-dimensional quantum axioms was carried out by Lucien Hardy~\cite{Hardy01a,Hardy01b} in 2001. In the subsequent years, numerous other reconstructions of the finite-dimensional axioms followed~(\eg \cite{Goyal-QT2c,GKS-PRA, DakicBrukner2010, Goyal2014, Hohn2017, Dariano-operational-axioms,Chiribella2011, MullerMasanes2016, MasanesMullerAugusiakPerez-Garcial2013,SelbyScandoloCoecke2018}), which approached the reconstructive challenge using a wide variety of mathematical and conceptual tools.

To elaborate, a reconstructor of quantum theory wishes to understand in readily graspable physical terms:
\begin{enumerate} 
\item \emph{von Neumann--Dirac axioms:}~Why quantum state space is complex~(as opposed to, say, real or quaternionic); why physical evolutions are represented by unitary transformations~(and not any other one-to-one map over state space); why the Born rule connects outcome probabilities to states in its precise manner~(and not in any other way); why states of subsystems compose via a tensor product operation~(and not some other binary operator).
\item \emph{Quantum wave equations:}~Why the Schroedinger and Dirac equations have their particular mathematical form~(and not some other).
\item\emph{Identical particle formalism:}~Why one must (anti-)symmetrize wavefunctions when constructing models of identical particle systems.
\item\emph{Classical-quantum correspondence rules:} Why classical and quantum physics are related through various operator correspondence and commutation relationships.
\end{enumerate}
In order to isolate this formalism from its experimental and historical context of discovery, the reconstructor employs an operational framework which makes only minimal assumptions about the physical system of interest.  For example, reconstructions of the von Neumann--Dirac axioms typically employ the minimal operational framework described in Sec.~\ref{sec:operational-framework}.  However, the reconstructions of other parts of the formalism~(such as the identical particle formalism) usually require an enlargening of this minimal operational framework, for example to describe~(and possibly differentiate between) measurements of static properties~(such as mass and charge) and dynamic properties~(such as position).

Over the last twenty-five years, reconstructions of most of the above-mentioned parts of the quantum formalism have been obtained:~\begin{inlinelist} \item the finite-dimensional von Neumann--Dirac axioms, from a variety of angles~(\eg~\cite{Hardy01a,Hardy01b, Goyal-QT2c,GKS-PRA,Goyal2014, DakicBrukner2010, Hohn2017, MullerMasanes2016, MasanesMullerAugusiakPerez-Garcial2013, Dariano-operational-axioms,Chiribella2011}); \item the quantum wave equations~(e.g.~\cite{DarianoPerinotti2014,Goyal-correspondence-rules-2010}); \item the identical particle formalism~\cite{Goyal2015,Goyal2019a,SanchezDakic2024}; and \item the classical-quantum correspondence rules~\cite{Goyal-correspondence-rules-2010}. \end{inlinelist} 

In Sec.~\ref{sec:reconstructive-interpretation}, we will examine how these reconstructive results can assist in  systematizing the interpretation of quantum theory.

\section{Interpretation of Quantum Theory}
\label{sec:interpretation-of-quantum-theory}

\subsection{Types of Interpretation}
\label{sec:interpretation-types}

There is no consensus or standard definition of what an `interpretation' of a physical theory consists of~(see, for example,~\cite[p.~8]{Laloe2022}).  Rather, there exist a wide spectrum of views amongst physicists and philosophers of physics which loosely fall under this banner.  While recognizing that any classification is arbitrary, it is helpful to distinguish amongst the following three types of interpretation:
\begin{itemize}
\item \textbf{Type 1.} \label{item:minimal-physical-interpretation} \emph{Minimal Physical Interpretation.} A minimal understanding of quantum theory that an experimental or theoretical physicist must possess in order to design and conduct experiments, and to construct viable theoretical models of physical systems of interest. 
\item  \textbf{Type 2.}\label{item:minimal-philosophical-interpretation} \emph{Minimal Philosophical Interpretation.} An interpretation that seeks to reconcile one or more philosophical viewpoints with quantum theory by making a set of minimal claims about the nature of the physical world.  
\item  \textbf{Type 3.}  \label{item:metaphysical-interpretation} \emph{Metaphysical Interpretation.} An interpretation that seeks a metaphysical understanding of part of the physical world described by quantum theory.
\end{itemize}
We shall refer henceforth to Types~2 and~3 as \emph{philosophical} interpretations.

The minimal physical interpretation incorporates a physicist's know-how in experimental design and modelling heuristics.  Some physicists would not describe this as an interpretation at all~(see \eg~\cite{FuchsPeres2000}).  However, such an interpretation usually employs assumptions and language which amounts to an informal metaphysics.  Some instances of such language, such as the frequently-encountered metaphor~(see \eg~\cite[p.~32]{Calosi2022}) that a diffracting electron passes through two slits at once, appears incidental.  However, in other instances, such as the interpretation of indices in multi-electron wavefunctions as \emph{particle} labels, the language importantly guides the application of the formalism~(see Difficulty 3, below), and so does not appear to be ignorable.

A minimal philosophical interpretation goes beyond what is strictly needed for~(at least) the \emph{routine} physical application of quantum theory, and seeks to give a schematic account of the physical world which reconciles quantum theory with some strongly-held views about reality or the nature of physical theory.  Some such interpretations are based directly on the standard formulation of quantum theory, whereas some are based on modifications thereof~(such as Bohm's hidden variable theory~\cite{Bohm52} or GRW's collapse model~\cite{GRW}).  For example, several well-known interpretations, such as the many worlds interpretation, are motivated by the universalist assumption~(Sec.~\ref{sec:bias3}) that unitary quantum theory applies to reality as a whole.

Finally, metaphysical interpretative work seeks to understand some part of the physical reality described by quantum theory in explicitly metaphysical terms.  Examples include the nature of identical quantum `particles'~(\eg~\cite{Schroedinger1950,DieksLubberdink2011,DieksLubberdink2020,Dieks2020,Bigaj2020b,French2000,FrenchKrause2006}), developing an understanding of the Bell-violating non-local correlations in causal terms~(\eg~\cite{IsmaelSchaffer2020}), and developing a conception of property adequate to describe quantum systems~(\eg~\cite{Margenau1954, Heisenberg1955, Heisenberg1958, Shimony1997, Karakostas2007, Suarez2007, Jaeger2017, Kastner2018}).   Such work typically sets up a contrast with parts of the classical conception of physical reality in light of quantum theory, and thereby elevates understanding of quantum theory by using classical physics as a point of contrast~(in the sense of van Fraassen's theory of questions and explanation~\cite{vanFraassen1980}).  Although such interpretative work is usually focussed on particular aspects of quantum theory, a \emph{complete} metaphysical interpretation would ideally provide a conception of reality of richness and coherence comparable to the classical conception sketched in Fig.~\ref{fig:classical-physics}, and would thereby enhance understanding of quantum theory by using the classical conception as a point of contrast.

\subsection{Interpretative Challenges and Difficulties}

The philosophical interpretation of quantum theory faces a number of obstacles.  Some of these are common to the interpretation of classical theories, as already discussed, but take on a more extreme form in quantum theory:
\begin{itemize}
\item  \label{item:challenge1} \textbf{Challenge 1. } \emph{Substantial physical content in mathematical terms.}  Most of the physical content of the quantum formalism~(e.g. the von Neumann--Dirac axioms) is encapsulated in mathematical relations.  Very little of its content is exposed in natural language.
\item  \label{item:challenge2} \textbf{Challenge 2. } \emph{Remoteness of quantum objects.} Unlike classical mechanics, quantum theory describes objects~(electrons, photons, and so on) which are not experimentally observed with the unaided senses.  Consequently, the experimental procedures for measurement of basic quantities such as location and mass may unwittingly project non-trivial assumptions about the nature of these objects\footnote{For example, as discussed in~\cite[\S2]{Goyal2023a}, in the measurement of time-independent properties of a quantum object such as an electron, it is implicitly assumed that the object is \emph{persistent}.  However, this assumption is in tension with a number of prevailing metaphysical assertions~(such as such objects being `non-individuals') about the nature of quantum objects in systems of identical particles.}.   In addition, experimental design itself is shaped by the quantum formalism in subtle ways.  
\item  \label{item:challenge3} \textbf{Challenge 3. } \emph{Complex modelling heuristics.} Construction of models of quantum systems involves complex heuristics which go beyond those needed for classical systems.  These heuristics involve the construction of classical foil models as intermediate modelling steps, and the application of rules of thumb to demarcate systems.
\end{itemize}
The impact of the interpretational biases discussed in Sec.~\ref{sec:interpretational-biases} is therefore intensified.
In addition, there are a number of special difficulties that hinder interpretation of the quantum formalism:
\begin{itemize} 
\item  \label{item:difficulty1}\textbf{Difficulty 1. }\emph{Metaphysically-laden natural language.}  The natural language component of the quantum formalism invokes numerous terms, such as \emph{state}, \emph{measurement}, and \emph{particle}, which, due to their use in classical physics, are unavoidably encumbered with a specific metaphysical meaning.  For example, \emph{state} evokes the standard metaphysical interpretation of the classical mechanical state, \emph{viz.} the \emph{objective physical state} of a body.  Similarly, \emph{measurement} evokes the notion of a process which \emph{passively registers} a pre-existing property.  None of these prior metaphysical associations are necessarily appropriate, but may colour ones understanding of the formalism since a firm alternative metaphysical association is not yet available.

\item  \label{item:difficulty2}  \textbf{Difficulty 2. }\emph{Justificatory or interpretative language.} Statements of certain parts of the formalism typically combine the bare-minimum that is needed for application of the theory~(and thus what is strictly empirically warranted) with additional concepts which serve in a justificatory or interpretative role.  For instance, the identical particle formalism is usually accompanied by the idea that identical particles are `indistinguishable', a term used to justify and interpret the identical particle formalism which can be traced to Dirac and Heisenberg~\cite{Dirac1926,Heisenberg1926}.  However, the correctness of this term is questionable~(see e.g.~\cite[\S1]{Goyal2019a}, and \S\ref{sec:reconstructive-interpretation} below). 

\item   \label{item:difficulty3} \textbf{Difficulty 3. }\emph{Non-trivial interpretation of symbolism.} The mathematical symbolism of certain parts of the quantum formalism is habitually interpreted in specific ways.  For example, the indices in symmetrized states of identical particle systems are almost universally interpreted as \emph{particle labels.}  However, this interpretative assumption has been recently brought into question~\cite{CaultonPhDthesis,DieksLubberdink2011,Goyal2019a}.
\end{itemize}
For these reasons, even the natural language which habitually accompanies the mathematical component of the quantum formalism can be misleading.

Below, we first discuss the main widely-used approaches to interpreting quantum theory in terms of the above-mentioned challenges and difficulties, and in terms of interpretational biases.  We then discuss the reconstructive approach to interpretation.

\subsection{Standard Interpretational Methods}
There are two widely-used methods for the philosophical interpretation of quantum theory.  The first constructs an interpretation on the basis of some part of the quantum formalism~(typically the von Neumann axioms, the Schr\"odinger equation, or the identical particle formalism) or on the basis of a modification thereof~(for example, Bohm's hidden variable theory).  For reasons that will become clear below, I shall refer to this method as the \emph{schematic interpretative method.}  The second is based not on a part of the quantum formalism \emph{per se}, but on a no-go theorem, such as the Bell or Kochen--Specker theorem---hence the \emph{no-go interpretative method.}  These two methods account for the most widely discussed interpretations of Type 2~(such as the \MWns,  \dBBns, and  \LB interpretations, together with the others of this type discussed in~\cite[pt.~V]{Freire2022}), and most interpretative work of Type 3~(such as investigations into the nature of identical quantum particles).

\subsubsection{Schematic interpretations of quantum theory}
Standard interpretations of the von Neumann axioms manifest all three of the previously-detailed interpretational biases~(Sec.~\ref{sec:interpretational-biases}).

\bigskip
\paragraph{Manifestation of Bias 1: Natural language over mathematical details.}
\subparagraph{Standard interpretations of von Neumann axioms.}  
\label{sec:standard-von-Neumann-interpretation}
Interpretations ranging from the \MW interpretation to the \LB interpretation, although diverse in many respects, are united by the common fact that they do not take into account the detailed mathematical structure of the von Neumann axioms.  Given Challenge~1, this is their greatest up-front limitation. 

\begin{figure}
\begin{center}
\includegraphics[width=\textwidth]{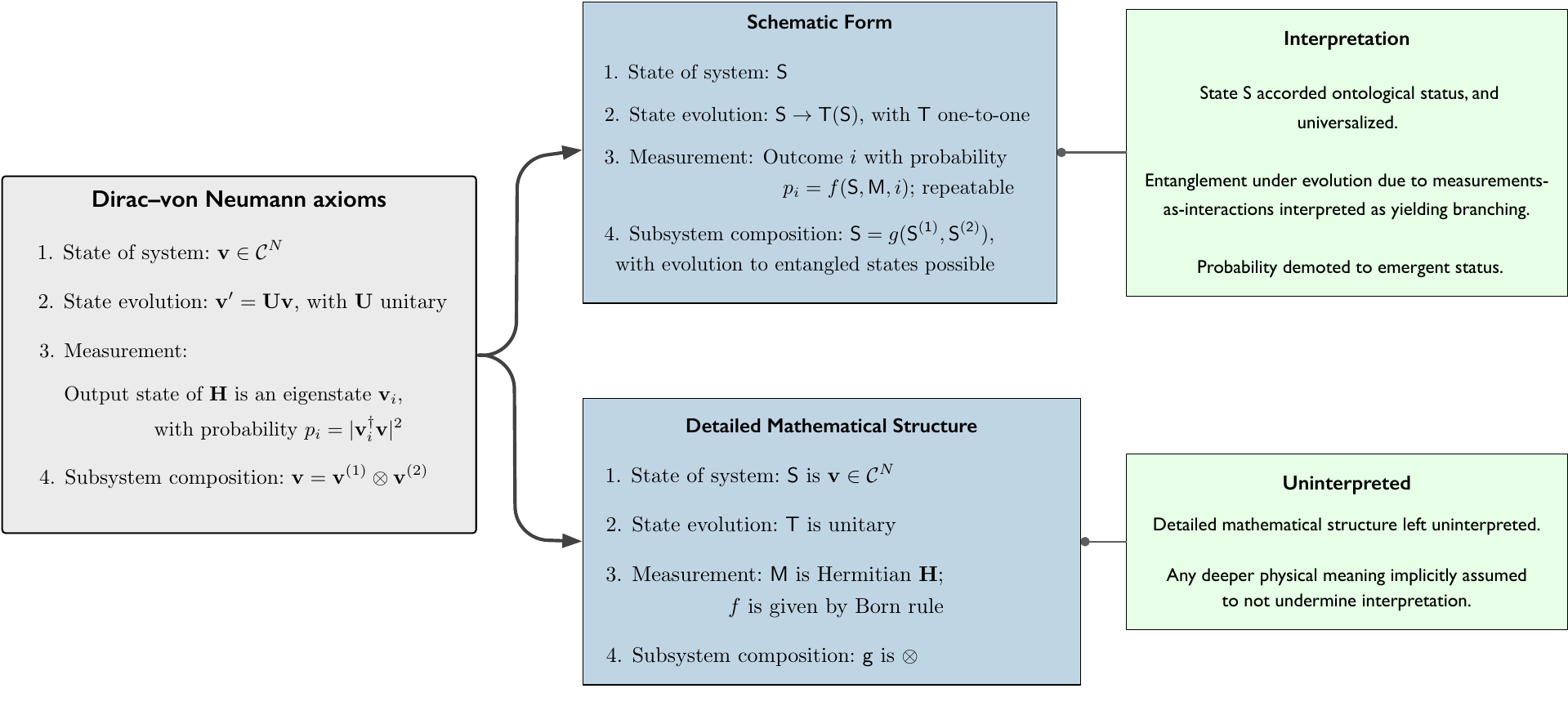}
\caption{\label{fig:schematic-interpretation}\emph{Schematic interpretation of the von Neumann axioms in the \MW interpretation.} }
\end{center}
\end{figure}

For example, the fact that quantum states are represented by complex vectors in a Hilbert space is not reflected in a typical interpretation.  Rather, the interpretation reflects only a \emph{schematic form} of the axioms.  This consists of what can be readily extracted by inspection of the natural language component of the axioms, together with some additional well-known logical implications of the axioms which can be expressed in natural language.  For example, the \MW interpretation additionally depends upon the existence of entangled~(\ie non-factorizable) states~(see Fig.~\ref{fig:schematic-interpretation}).

This dependency on natural language offers little protection against Difficulty~1.  The appeal to entangled states raises this difficulty further:~a bipartite quantum state is said to consist of two subsystems.  This technical language evokes the metaphysical notion that these two subsystems are \emph{individuals}, akin to the standard interpretation of the atoms or particles of classical mechanics.  However, the existence of entangled states brings into question the applicability of the metaphysical notion of individual~(along with the associated substance--property dichotomy).

\begin{figure}
\begin{center}
\includegraphics[width=\textwidth]{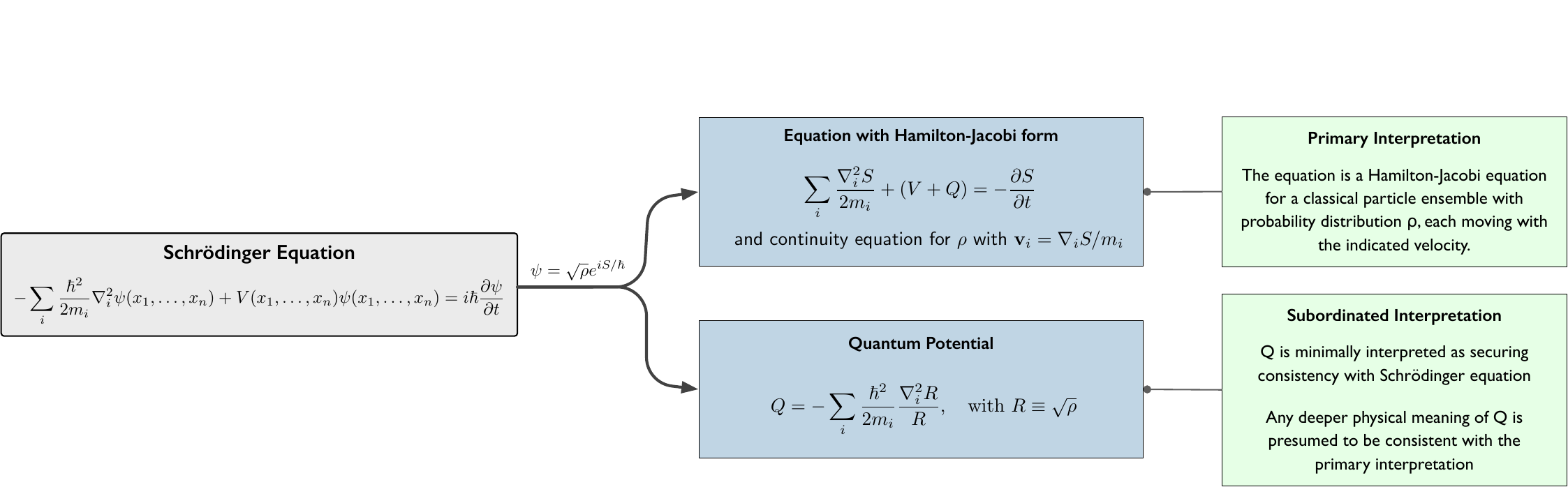}
\caption{\label{fig:Bohm-interpretation}\emph{Bohmian interpretation of the Schr\"odinger equation.}}
\end{center}
\end{figure}

\subparagraph{Standard interpretations of the Schr\"odinger equation.} 
\label{sec:standard-Schroedinger-interpretation}
A similar situation obtains with interpretations of the Schr\"odinger equation, where `natural language' is expanded to include equations familiar from classical mechanics.  For example, the \dBB interpretation depends on a rewriting of the Schr\"odinger equation in a form which is symbolically isomorphic to the Hamilton-Jacobi equation~(see Fig.~\ref{fig:Bohm-interpretation}).
The interpretation is then based on the Hamilton-Jacobi look-alike.  This suggests an interpretation based on a classical particle ontology, which is then made explicit by positing the existence of unobserved point particles and an associated guidance equation taken from classical physics.  The interpretation of the specific mathematical form of the potential term is then subordinated to this interpretation.  That is, the mathematical form of the so-called quantum potential is interpreted as if it were simply part of the potential in a classical Hamilton-Jacobi equation.  But that interpretation does not reflect the \emph{precise mathematical form} of the potential---the interpretation would still stand if the potential term were different. 

This interpretational pattern is analogous to the aether interpretation of Maxwell's equations.  As described in~Sec.~\ref{sec:Maxwells-equations}, in that case the interpretation is based on a mechanical interpretation of electromagnetic waves~(which leads to the posit of an unobserved aether), while the Lorentz transformations are primarily regarded as means to ensure non-detectability of motion relative to the aether.  The insensitivity of the aether interpretation to the precise mathematical form of Maxwell's equations is evident---the interpretation would hold even under considerable deformation of Maxwell's equations and the associated transformations, provided the equations were to yield wave-like solutions.

\subparagraph{Summary.}
The standard interpretations are schematic---they only reflect a limited natural-language~(or classical look-alike)  fraction of the quantum formalism~(whether the von Neumann axioms or the Schr\"odinger equation).  They either do not interpret the precise mathematical content, or subordinate its interpretation to an interpretation of the schematic part.  These interpretations implicitly assume that the elucidation of the physical meaning of the exact mathematical form~(of the von Neumann axioms, or the Schr\"odinger equation) will not undermine the interpretation.  However, as we have seen in the example of the aether interpretation of Maxwell's equations, this is an unsafe assumption.  
\bigskip
\paragraph{Manifestation of Biases 2 \& 3.  Formalism over experimental practice and modelling heuristics.}
\label{sec:standard-interpretation-bias23}
The standard interpretations make limited and uneven use of the informal part of the theory~(\emph{viz.} the operational framework and its experimental practices; and modelling heuristics).  In some cases, the informal part is ignored altogether; in other cases, it is used selectively.  

For example, the \MW interpretation presumes that a quantum system can encompass the universe as a whole.  The \LB interpretation is similar, except that it exempts the consciousness of the observer. However, as discussed in Sec.~\ref{sec:bias3}, such universalizing assumptions extrapolate quantum theory far beyond its demonstrated domain of empirical validity.

\subsubsection{Interpretations based on no-go theorems}
A no-go theorem, such as the Bell or Kochen--Specker theorem, employs a minimal operational framework~(subject to operationally-grounded assumptions such as no-conspiracy), and then formalizes within an operational language particular constraints inspired by classical physics, which yields a mathematical condition expressed in terms of measurement outcome probabilities.  The result of the theorem is that this mathematical condition is violated by certain predictions of quantum theory. 

An \emph{interpretation} of a no-go result seeks to develop a philosophical understanding of quantum objects which renders the no-go result intelligible.
Such an interpretational approach avoids Challenge 1 insofar as the interpretationally relevant mathematical content has been distilled into an explicit no-go result whose meaning is \emph{operationally} unambiguous.  However, \emph{metaphysical} interpretation of the result is still subject to Difficulties~1 and~3.  For example, in the interpretation of the violation of Bell's inequality, it is commonplace to speak of subsystems as if they were individuals, despite the existence of entangled states~(as already mentioned in Sec.~\ref{sec:standard-von-Neumann-interpretation} above).  Unless this formalism is securely connected to the operational framework through specific experimental practices~(\ie countering Bias~2), the appropriate way to think metaphysically about these subsystems will remain ambiguous.

The broader issue here is that no-go theorems are inherently \emph{negative} or \emph{eliminative}---they rule out a particular classical way of metaphysically framing a given situation~(\eg in Bell's theorem, of thinking of two widely-separated particles as individuals interacting through local interactions), but they do not provide a clear \emph{positive} path forward that leads to a new way of thinking about the situation.  Any attempt to develop a positive understanding that engages with the quantum formalism will bring the above-mentioned Challenges and Difficulties back into play, despite the fact that the no-go theorem itself is free of such issues.

A final limitation of interpretations based on no-go theorems is that these theorems are \emph{fragmentary}---they bring specific non-classical aspects of quantum theory into sharp focus, but the integration of the interpretations of several such fragmentary no-go results is exceedingly challenging.  Moreover, there are no known no-go results for many aspects of quantum theory~(such as the identical particle formalism).  Hence, one cannot hope to build a comprehensive interpretation of quantum theory on the basis of a variety of no-go results.

\subsection{Reconstructive Interpretation of Quantum Theory}
\label{sec:reconstructive-interpretation}

A reconstruction-based interpretation tackles many of the challenges and difficulties delineated in Sec.~\ref{sec:interpretation-of-quantum-theory}.  First, reconstruction of a part of the quantum formalism distills the entire content of that formalism into a set of postulates.  These postulates will, in general, contain both a natural language and mathematical component.  Interpretation of these postulates rather than the quantum formalism is beneficial in several ways:
\begin{enumerate} 
\item\emph{Operational meaning of natural language.}  The natural language will~(at least ideally) have an unambiguous operational meaning.  For example, in reconstructions of the von Neumann axioms,  `state' is simply a mathematical object which is introduced in order to organize measurement outcome probabilities for different measurements.  No additional metaphysical meaning is implied.  \emph{(Resolves Difficulty 1).}
\item\emph{Simple mathematical component.}  Any mathematical component will be considerably simplified in comparison to the quantum formalism.  For example, in Hardy's reconstruction~\cite{Hardy01a}, the postulates contain only elementary expressions involving the product of integers.  There are no complex vector spaces, unitary operators, \etc \emph{(Ameliorates Challenge 1, and  Redresses Bias 1).}
\item\emph{Intelligible principles.}  The postulates will usually be graspable as expressions of \emph{physical principles}.  In an ideal reconstruction, all of the postulates can be viewed as expressions of intelligible principles.  For example, a common postulate in the reconstruction of the von Neumann axioms is \emph{local tomography}, which can be viewed as a probabilistic expression of the desideratum of \emph{local knowability}, \emph{viz.} the possibility of learning about the whole by combining locally-gained knowledge. \emph{(Ameliorates Challenge 1, Redresses Bias 1).}
\item\emph{No justificatory or interpretative language.} Since the formalism is being derived \emph{de novo}, there is no confounding historical legacy of justificatory or interpretative language.  For example, in the reconstruction of the identical particle formalism in Refs.~\cite{Goyal2015, Goyal2019a}, the formalism is derived free of the widespread justificatory and interpretative notion that identical particles are `indistinguishable'.   \emph{(Resolves Difficulty 2).}
\item\emph{Unambiguous interpretation of symbolism.} Since the formalism is derived in the context of an operational framework, the interpretation of the symbolism is operationally unambiguous.  For example, the labels in a tensor-product state refer to subsystems which the operational framework presumes can be distinguished through experimental procedures.  Similarly, Refs.~\cite{Goyal2015, Goyal2019a} provide clear operational meaning to the indices in symmetrized states of identical particle systems.  \emph{(Resolves Difficulty 3).}
\end{enumerate}
   
Second, since the formalism is now transparently derived from a clear operational basis, it becomes much easier to see where it requires elucidation.  For example, in tensor-product states, the labels refer to operationally-distinguishable subsystems.  But that provokes the question: what are the experimental procedures that actually enable such distinguishing, and what are the assumptions implicit in these procedures?  This naturally leads to further issues such as the primacy of spatial measurements, the measurement of static properties in terms of spatial measurements, persistence and reidentification, and so on.   This directly redresses Bias 2.

Third, the derivation of the formalism within an operational framework keeps the distinction between~(at minimum) system, environment, measurement, and observation, constantly in view.  This counters the universalising tendencies discussed above~(\S\ref{sec:standard-interpretation-bias23}), and redirects attention towards the experimental procedures which substantiate these distinctions.

Reconstruction-based interpretation also offers a number of special advantages.  First, the opportunity to philosophically reflect on physical principles articulated or even discovered during the process of reconstruction.  Second, the possibility of \emph{discovering} new physical principles through the act of interpreting a reconstruction.

Many of the principles used in a reconstruction are formalisations of facts that can be deduced from the quantum formalism.  Examples include the above-mentioned principle of local tomography, and the requirement of symmetric transition probabilities~\cite{GKS-PRA}.  Although these principles may seem mundane, they can have remarkable consequences.  For instance, local tomography is a base assumption in so-called generalized probabilistic theories, and many remarkable consequences have been shown to flow from this apparently simple assumption~(\eg~\cite{Barrett2006}).  In this connection, compare Einstein's equivalence principle, which was directly inspired by a mundane-seeming fact that follows from Newtonian mechanics, a fact that Darrigol has termed the principle of accelerative relativity~\cite{Darrigol2019}.

However, interpretation of reconstructions can sometimes lead to discovery of principles that are genuinely unfamiliar.  For example, a reconstruction of the identical particle formalism~\cite{Goyal2015} is based on the Operational Indistinguishability Postulate~(OIP).  Interpretation of this reconstruction has given rise to a new principle of complementarity, the `Complementarity of Persistence and Nonpersistence'~\cite{Goyal2019a}, which has already attracted philosophical interpretative interest~\cite{Jantzen2020,Bitbol2023}.

In addition, we anticipate that interpretation of extant reconstructions will give new, precise formulations of vague yet intriguing notions promoted by the founders, such as Bohr's complementarity, and Heisenberg's notions of potentiality and actuality.  In this way, reconstructive interpretation can widen the range of philosophical viewpoints and traditions which can creatively engage with quantum theory.

\section{Concluding remarks}
\label{sec:conclusion}

Over the past 25 years, the quantum reconstruction program has yielded a rich trove of reconstructive results which, as late as the 1990s, seemed out of reach.  However, apart from a couple of isolated exceptions, these results have yet to be used for substantive interpretative purposes.  In this paper, we have shown that these reconstructive results offer a new pathway to the interpretation of quantum theory.  In particular, we have shown how they can help systematize the interpretational process by allowing an interpreter to take into account more of the potentially relevant content of quantum theory---both content encoded in the quantum formalism and content implicit in its experimental procedures and modelling heuristics.  Furthermore, as this methodology makes available more of the content of quantum theory for philosophical reflection, more detailed interpretations~(particularly those of Type~3) are brought within reach.

The \emph{implementation} of this reconstruction-based methodology involves a number of challenging aspects.  In particular, the success of the methodology depends upon judicious selection of reconstructions.  As there are now numerous reconstructions of the core quantum formalism, it will be necessary to develop experience in identifying those that are well-suited for interpretative purposes.  Such experience will likely accumulate through a body of reconstruction-based interpretative case studies, followed by the formulation of a set of heuristics.   In addition, as mentioned in Sec.~\ref{sec:reconstructive-interpretation}, the now-extensive work in the area of generalized probabilistic theories furnish what may be regarded as \emph{partial} reconstructions, the interpretation of which provides an opportunity to investigate generalized theoretical structures of which quantum theory is a special case.

Another challenge to implementation of this methodology is the academic compartmentalization of those who have carried out reconstructive work and those who are primarily engaged in the interpretation of quantum theory.  We hope that this paper contributes to the bridging of these compartments.

\begin{acknowledgements}
I am grateful for the constructive feedback received during a workshop at Link\"oping University~(organized by Harald Wiltsche and Philipp Berghofer) at which many of the ideas contained herein were presented.  In particular, I express my thanks to Steven French, Philipp Berghofer, Harald Wiltsche, Tom Ryckman, Daniele Pizzocaro, Sebastian Horvat, Luis Barbado, Jessica Oddan, and Lars-G\"oran Johansson for insightful remarks, and the anonymous reviewers for their thoughtful and insightful comments.
\end{acknowledgements}

\bibliography{references_master}
\end{document}